\documentclass[lineno]{jfm}

\usepackage{graphicx}
\usepackage{newtxtext}
\usepackage{newtxmath}
\usepackage{natbib}
\usepackage{hyperref}
\hypersetup{
    colorlinks = true,
    urlcolor   = blue,
    citecolor  = black,
}
\usepackage{array} % for better arrays (eg matrices) in maths
\usepackage{paralist} % very flexible & customisable lists (eg. enumerate/itemize, etc.)
\usepackage{verbatim} % adds environment for commenting out blocks of text & for better verbatim
\usepackage{subfigure} % make it possible to include more than one captioned figure/table in a single float
\usepackage[normalem]{ulem}
\usepackage{physics}
\def\be{\begin{equation}}
\def\ee{\end{equation}}
\def\bea{\begin{eqnarray}}
\def\eea{\end{eqnarray}}

\graphicspath{{figures_m1/}}
\begin{document}
\title{Birth, interactions, and  evolution over topography of solitons in Serre-Green-Naghdi model}

\author{Qingcheng Fu\aff{1}, Alexander Kurganov\aff{1}\aff{2}\aff{3}, Mingye Na\aff{1} \and Vladimir Zeitlin\aff{4} \aff{2}
\corresp{\email{zeitlin@lmd.ens.fr}}} 

\affiliation{\aff{1}Department of Mathematics, Southern University of Science and Technology, Shenzhen, 518055, China
\aff{2} Shenzhen International Center for Mathematics, Southern University of Science and Technology, Shenzhen, 518055, China
\aff{3} Guangdong Provincial Key Laboratory of Computational Science and Material Design, Southern University of Science and Technology,
Shenzhen, 518055, China
\aff{4}Laboratory of Dynamical Meteorology, Sorbonne University, Ecole Normale Sup\'erieure, CNRS, 75005 Paris, France}
\maketitle

\begin{abstract}
New evidence of surprising robustness of solitary-wave solutions of the Serre-Green-Naghdi (SGN) equations is presented on the basis of
high-resolution numerical simulations conducted using a novel well-balanced finite-volume method. SGN solitons exhibit  a striking
resemblance with their celebrated Korteweg-deVries (KdV) counterparts. Co-moving solitons are shown to exit intact from double and triple
collisions with a remarkably small wave-wake residual. The counter-propagating solitons experiencing frontal collisions and solitons hitting a wall,  non-existing in KdV case configurations,  are shown to also recover, but with a much larger than in co-moving case residual, confirming with higher precision the results known in the literature.  Multiple SGN solitons  emerging from localized initial conditions are exhibited, and  it is demonstrated that SGN solitons
survive hitting  localized topographic obstacles, and generate secondary solitons when they encounter a rising escarpment.  
\end{abstract}


%{\bf MSC Codes }  {\it(Optional)} Please enter your MSC Codes here
\section{Introduction}
\label{sec: intro}
{Serre-Green-Naghdi (SGN) equations is a  generalization of the classical shallow-water equations obtained by relaxing the hydrostatic approximation in the standard
derivation, cf. e.g. \citep{DellarSalmon}. The model was first derived by \cite{Serre} and then rediscovered by \cite{SuGardner} in
the one-dimensional configuration relevant to the present study, although without topography. The emphasis in the later work
\cite{GreenNaghdi} was on two-dimensional motions in the presence of topography.}  
Finite-amplitude steady-moving solutions in a form of periodic cnoidal waves, and solitary waves as their limiting form, are known in the
SGN system starting from \citep{Serre, SuGardner}. It is also well-known \citep{SuGardner}, that the celebrated Korteweg-deVries (KdV)
equation arises as an asymptotic limit of the SGN equations in the case of unidirectional weakly nonlinear waves. The KdV equation possesses
exact solutions in the form of cnoidal waves and solitons, which were known for a long time. Yet one of the most striking properties of the
KdV equation is that it possesses also exact multi-soliton solutions. (We will place ourselves from now on in the framework of decaying
boundary conditions, and will not discuss periodic waves). These multi-soliton  solutions were first discovered numerically in a landmark
paper by \cite{ZabuskyKruskal}, and their existence triggered the studies of complete integrability of the KdV equation, which was later
successfully proved. In fact, any localized initial condition for the KdV equation is being transformed into a sequence of solitons of
different amplitudes and speeds. Let us recall that Zabusky and Kruskal observed a striking behavior of the  KdV solitons, which they used to initialize numerical simulations: the solitons passed one through another recovering their initial shape after multiple collisions.

A natural  idea  to make numerical experiments {\em\`a la} \cite{ZabuskyKruskal} with SGN solitons, in order to compare the behavior of
corresponding multi-soliton configurations, arises in this context. The SGN solitons are routinely  used for testing numerical methods for
the SGN equations, e.g. \citep{Metayer,PearceEsler,BonnetonETAL}, while existing studies of their collisions are sparse
\citep{MirieSu,DutykhETAL,MitsotakisETAL}, limited to collisions of pairs of SGN solitons, and mostly in the frontal configuration which is
motivated by experimental studies, e.g. \citep{ChengYeh}, and is impossible for KdV solitons. A study in the aforementioned spirit is by
\cite{CraigETAL}, where strong resemblance with the behavior of KdV solitons was emphasized, but still on the basis of pairwise soliton
collisions, and in the framework of the full Euler equations (see a corpus of literature on theoretical, experimental, and numerical
investigations on collisions of soliton solutions of Euler equations therein)---not for the SGN model. On the other hand, laboratory
experiments on soliton interactions with localized topographic obstacles, and early numerical simulations  \citep{Seabra-SantosETAL}  show that solitons survive collisions with bumps and escarpments, and produce a secondary trailing soliton-like structure in the latter case  

In this paper, we present numerical experiments on collisions of SGN solitons and their interactions with topography. The results obtained
with the help of a novel well-balanced high-resolution numerical method, confirm the surprising robustness of SGN solitons. We study not
only double, but also triple-soliton configurations, and show that co-moving SGN solitons basically reproduce the behavior of KdV solitons.
The SGN solitons, unlike the KdV ones, can propagate in both directions, so we also study frontal collisions, and collisions with
a vertical wall. We also present results of experiments on interaction of SGN solitons with topographic obstacles, giving additional
evidence of their robustness. Finally, we show that localized initial conditions give rise to one- or multi-soliton  configurations, depending on their shape. The bulk of our results supports the conjecture that  multi-soliton configurations are
attracting quasi-exact solutions of the SGN equations. 

\section{SGN equations, their soliton solutions, and {key points} of the numerical method}\label{sec: model-and-solitons}
The SGN equations in one spatial dimension $x$ read as
\begin{equation}
\left\{\begin{aligned} 
&h_t+uh_x+hu_x=0,\\
&u_t+uu_x+g(h+b)_x+\frac{1}{3h}\left(h^2{\cal D}^2\Big(h+{3\over2}b\Big)\right)_x+b_x{\cal D}^2\Big({1\over2}h+b\Big)=0, 
\end{aligned}\right.
\label{2.1}
\end{equation}
where $h(x,t)$ is the thickness of the fluid layer, $u(x,t)$ is its velocity, $b(x)$ is the bottom topography, {$g$ is the
acceleration due to gravity, and ${\cal D}:=\frac{\partial}{\partial t}+u\frac{\partial}{\partial x}$} is the material time derivative.
Localized exact solutions of \eqref{2.1} with $b\equiv0$, solitary waves or solitons, having a maximal amplitude $h_{\max}$, and moving with
a constant speed $c$ on the background of unperturbed depth $h_\infty$, are given by (e.g. \citep{Metayer}):
\begin{equation}
\left\{\begin{aligned}
&h(x,t)=\widehat h(x-ct)=h_\infty+(h_{\max}-h_\infty)\sech^2\left(\sqrt{\frac{3(h_{\max}-h_\infty)}{h_{\max}h^2_\infty}}\,\frac{(x-ct)}{2}
\right),\\ 
&u(x,t)=\widehat u(x-ct)=c\bigg(1-\frac{h_\infty}{\widehat h(x-ct)}\bigg),\quad c=\pm\sqrt{h_{\max}}.
\end{aligned}\right.
\label{2.2}
\end{equation}
These solutions can propagate with the same {speed $\sqrt{h_{\max}}$} both left- and rightwards. It is important
that they decay exponentially, so if several solitons are {initially} placed sufficiently far from each other, the overlapping of
their tails is exponentially small. Such configurations will be used for initialization in \S\ref{sec3}.

The simulations presented in \S\ref{sec3} have been conducted using a novel flux globalization based well-balanced central-upwind scheme. We
only sketch the main idea of its construction here, the full details will be presented elsewhere. We first reformulate the system
\eqref{2.1} in the following quasi-conservative form:
\begin{equation*}
\left\{\begin{aligned}
&h_t+q_x=0,\\
&Q_t+K_x=0,
\end{aligned}\right.
\end{equation*}
where $q:=hu$, $Q:=q(1+h_xb_x+\frac{1}{2}hb_{xx}+b_x^2)-\frac{1}{3}(h^3u_x)_x$, and
$K$ is a global flux:% given by
\begin{equation*}
K:=hu^2+\frac{g}{2}h^2-\frac{h}{3}\big[hq_xu_x+(qhu_x)_x\big]+\frac{1}{2}(q^2b_x)_x+qub^2_x+
\int\limits^x\Big[ghb_x+qb_{xx}\Big(\frac{1}{2}hu_x-ub_x\Big)\Big]{\rm d}x.
\end{equation*}
We then design a well-balanced scheme, which is capable of exactly preserving certain physically relevant steady states satisfying
\begin{equation}
q\equiv{\rm Const},\quad K\equiv{\rm Const},
\label{2.3}
\end{equation}
at the discrete level. Note that formula \eqref{2.3} contains both ``moving-water'' $q\not\equiv0$ and ``still-water'' $q\equiv0$ steady
states \eqref{2.3}.

We construct our well-balanced scheme along the lines of \citep{ChertockCKOT,ChertockHO,ChenCHKW,KurganovLZ,ChertockKLLW}: we reconstruct
the equilibrium quantities $q$ and $K$ instead of $h$ and $q$, {and then recover the point values of $h$ by solving the corresponding
nonlinear equations. In additions, as the evolved quantities are $h$ and $Q$, we compute $q$ at the end of each time evolution step by
solving the corresponding tridiagonal linear systems. In order to reduce the numerical dissipatio present in the central-upwind scheme and to increase the efficiency of the resulting scheme, we implement the moving framework approach from \citep{KliKur}.}

\begin{figure}
\centerline{\includegraphics[width=0.5\columnwidth, height=4cm]{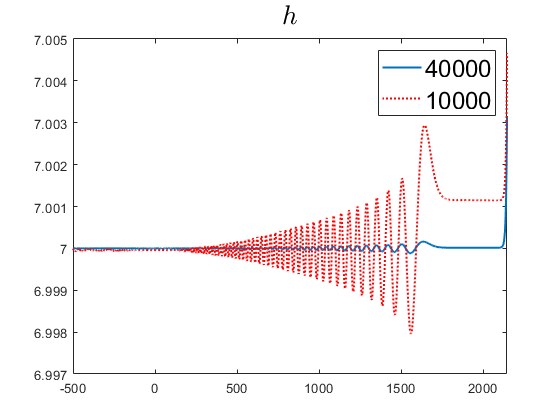}
            \includegraphics[width=0.5\columnwidth,height=4cm]{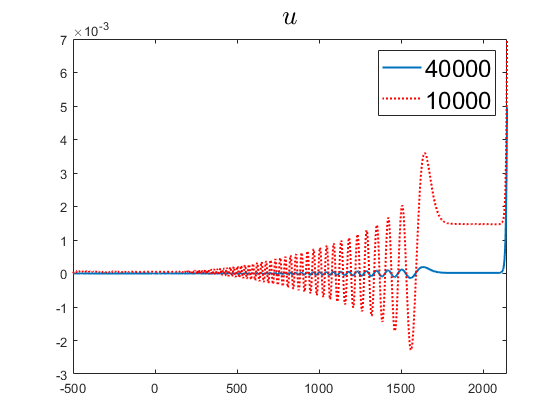}}
\caption{Zoom of umerical oscillations in $h$ (left) and $u$ (right) at the rear of the soliton at two resolutions with 40000 and 10000 cells in the computational domain $[-2500,2500]$.\label{fig1}}
\end{figure}

We emphasize that the soliton solution \eqref{2.2} in the reference frame moving with its speed is a {``still-water'' steady state}, and the numerical scheme is capable of {\em exactly}
preserve it provided the initial conditions are ``well-prepared''. Such preparation, however, requires solving for $h$ at given
{$q$ and $K$}, which is rather complicated and will not be done below. We, however, take advantage of the fact that the
well-balanced {property} of the scheme guarantees {that small numerical oscillations due to the
initial discretization errors do not amplified in time and also rapidly decay when the mesh is refined,} which we have checked. If the code
is initialized with \eqref{2.2}, the soliton keeps moving {while maintaining its initial shape} for as long as we follow it (not
shown), but zooming reveals small-amplitude oscillations at its rear. They are, however, rapidly decreasing with increasing resolution, as
shown in Fig. \ref{fig1}. The amplitude of these oscillations should be compared with the amplitude of the soliton $h_{\max}=12.1$.
We also stress that while the total energy of the system is not perfectly conserved during the
simulations, a typical overall decrease of the total energy normalized by its initial value is $\sim10^{-6}$.

\section{Results of numerical simulations}\label{sec3}
\subsection{Soliton collisions}
\subsubsection{Collisions of co-moving solitons}
\begin{figure}
    \begin{center}
   % \noindent\makebox[\columnwidth]{%
    \includegraphics[width=0.5\columnwidth, height=4cm
    ]{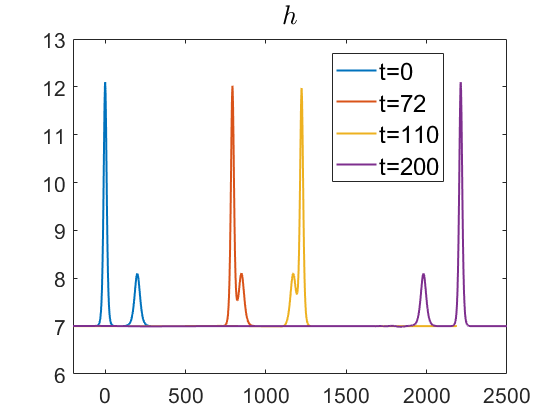}%\\
    \includegraphics[width=0.5\columnwidth, height=4cm
    ]{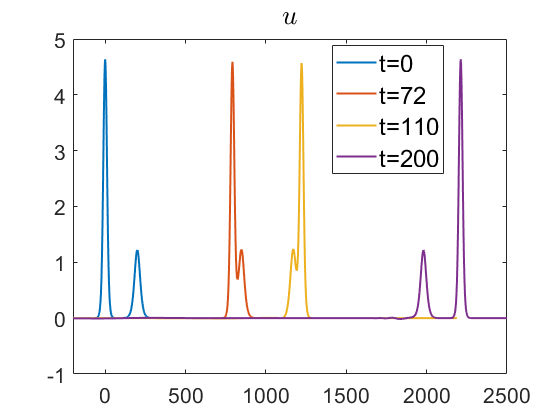}
%    }
   \end{center}
    \caption{Snapshots of a collision of two co-moving solitons with maximum heights, respectively $h_{max} = 12.1, 8.1$ for two solitons at  $H =7$. Left (right) panel: height (velocity) field.
       \label{fig: 2waves}
        }
\end{figure}

\begin{figure}
    \begin{center}
   % \noindent\makebox[\columnwidth]{%
    \includegraphics[width=0.5\columnwidth, height=4cm
    ]{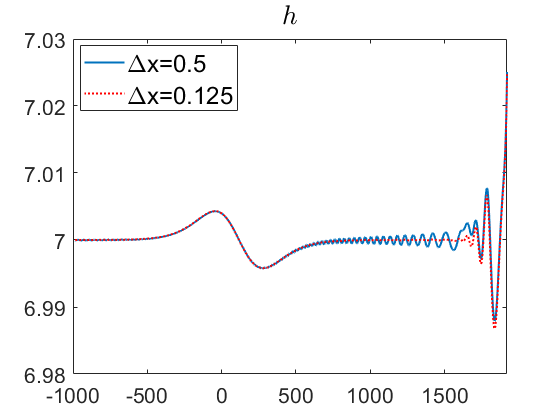}%\\
    \includegraphics[width=0.5\columnwidth, height=4cm
    ]{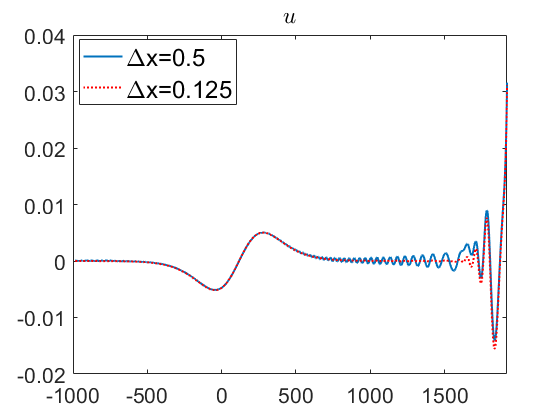}
%    }
   \end{center}
    \caption{Wave-wake produced by collision of a pair of co-moving solitons at two different resolutions.  Left (right) panel: height (velocity) field.
       \label{fig: wavetail2}
        }
\end{figure}

\begin{figure}
    \begin{center}
 \includegraphics[width=0.5\columnwidth, height=4cm
    ]{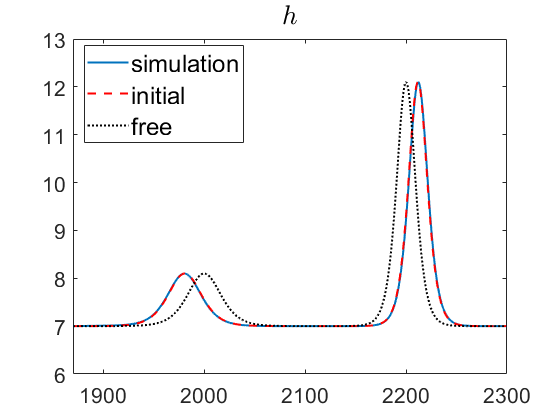}%\\
    \includegraphics[width=0.5\columnwidth, height=4cm
    ]{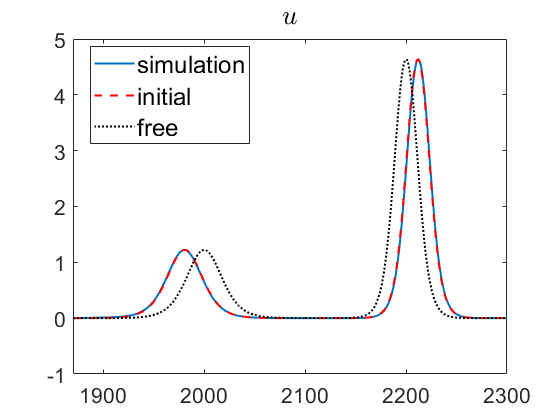} 
 \end{center}
    \caption{Initial data (dashed) for each soliton superimposed onto their final shapes (solid) at $t= 200$, and their respective positions in the case of free propagation with  the same time lag (dotted).  Left (right) panel: height (velocity) field.
       \label{fig: comparisons}
        }
\end{figure}

We start by a two-soliton encounter by placing a higher amplitude (faster) soliton behind a lower amplitude (slower) soliton, with the same sign of velocity, and at a sufficient distance, in order to have a negligible overlapping, and let them go. As follows from   Fig. \ref{fig: 2waves}, the rear soliton passes through the front one and rapidly recovers its form after collision. The collision produces a  rapidly dispersed wave packet of small amplitude gravity waves left behind the solitons. 
Both solitons engaged  in the collision preserve their form quasi-exactly, as follows from Fig. \ref{fig: comparisons}. The amplitude difference, not visible at the scale of the figure, is less than half per-cent point-wise, both in height and velocity (not shown).  The $L2$ norms of initial and final state are, respectively, $21.6138$, and $21.6131$.  The residual represents a small-amplitude wave wake, which is not of numerical origin, as its amplitude  does not decrease with increasing resolution, cf. Fig. \ref{fig: wavetail2}.
The existence of the wave-wake was proven theoretically by an asymptotic analysis for the same process in the framework of full Euler equations, cf. \citep{Byatt-Smith} and references therein, so there is no doubt that it is of physical origin. 
During the collision process the slower soliton accelerates, while the faster soliton decelerates, as follows from the comparison of their positions at the end of the simulation with those each of them would  have at the same moment of time if propagating freely by itself.

\begin{figure}
    \begin{center}
   % \noindent\makebox[\columnwidth]{%
    \includegraphics[width=0.5\columnwidth, height=4cm
    ]{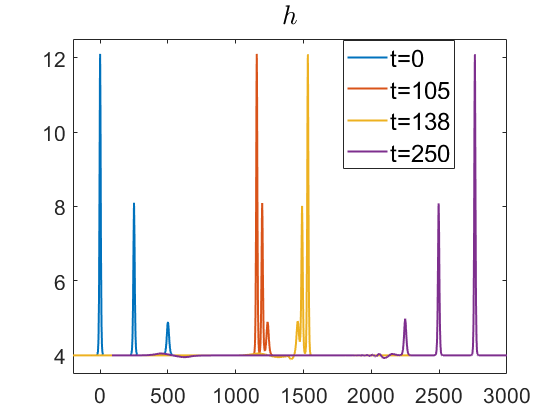}%\\
    \includegraphics[width=0.5\columnwidth, height=4cm
    ]{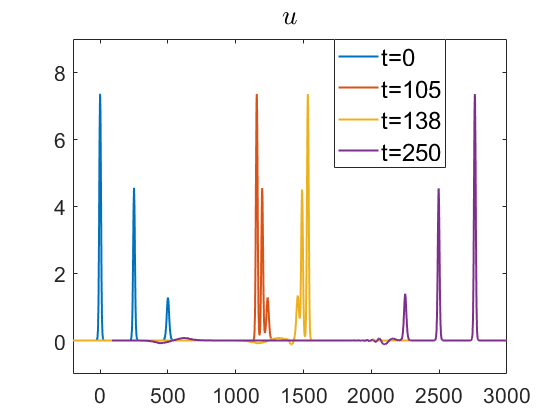}%\\
   \end{center}
    \caption{Triple collision of  three co-moving solitons with $h_{max} = 12.1, 8.1, 4.9$, respectively,  $H =4$.  Left (right) panel: height (velocity) field.
       \label{fig: 3wavesa} 
        }
\end{figure}

\begin{figure}
    \begin{center}
 \includegraphics[width=0.5\columnwidth, height=4cm
    ]{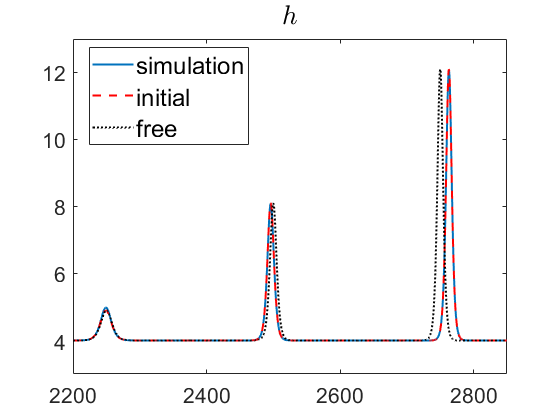}%\\
    \includegraphics[width=0.5\columnwidth, height=4cm
    ]{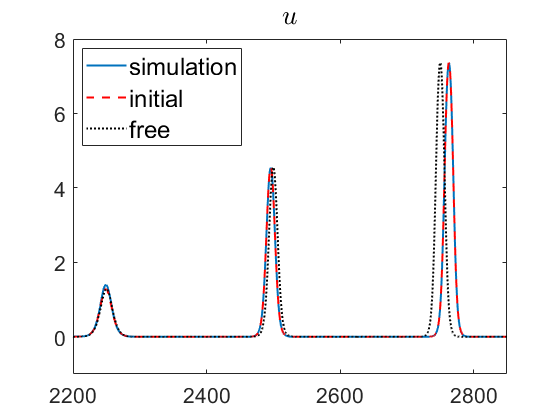} 
 \end{center}
    \caption{Initial data (dashed) for each soliton superimposed onto their calculated shapes (solid) at $t= 250$, and their respective positions in the case of free propagation with  the same time lag (dotted).  Left (right) panel: height (velocity) field.
       \label{fig: comparisons3}
        }
\end{figure}

We then examine three-soliton interactions. If we  place three solitons of increasing amplitude one behind another with a  lesser distance between the two front solitons, we observe sequential  pairwise collisions of the kind presented above. The solitons pass through each other recovering their form after collisions, which produce each time  emission of rapidly dispersing packets of low-amplitude gravity waves (not shown).  
%\begin{figure}
%    \begin{center}
   % \noindent\makebox[\columnwidth]{%
 %   \includegraphics[width=0.5\columnwidth%, height=9cm
 %   ]{3wavesh_b.png}%\\
  %  \includegraphics[width=0.5\columnwidth%, height=9cm
 %   ]{3wavesu_b.png}%\\
 %  \end{center}
%    \caption{Snapshots of a collision of three co-moving solitons with maximum heights, respectively \mingye {$h_{max} = 12.1, 8.1, 4.9$} for two solitons at  $H =4$. Left (right) panel: height (velocity) field.
%       \label{fig: 3wavesb} 
 %       }
%\end{figure}
We, however, can choose the parameters of the initial solitons in a way that they experience a simultaneous triple collision, coalescing at some moment. In this case, which to our knowledge was not reported in the literature,   the solitons also fully recover their initial form and steady motion, as follows from  Fig.  \ref{fig: 3wavesa},  The $L2$ norms of the system, respectively, are $25.5145$ at $t=0$,  and $25.5191$ at $t=250$, the quite negligible difference giving the idea of the smallness of magnitude of the wave-tail produced by the collision. 
%As in the previous examples we observe a small-amplitude wave-tail produced by the collision,. Increasing the resolution does not allow to get rid of it, as we checked. 
Similar to those of Fig. \ref{fig: comparisons} comparisons are presented in Fig. \ref{fig: comparisons3}, and show that the highest-amplitude soliton decelerates, while two other accelerate during the impact. 

\subsubsection{Collisions of counter-propagating solitons, and of a soliton with a wall}
\begin{figure}
    \begin{center}
   % \noindent\makebox[\columnwidth]{%
    \includegraphics[width=0.5\columnwidth%, height=9cm
    ]{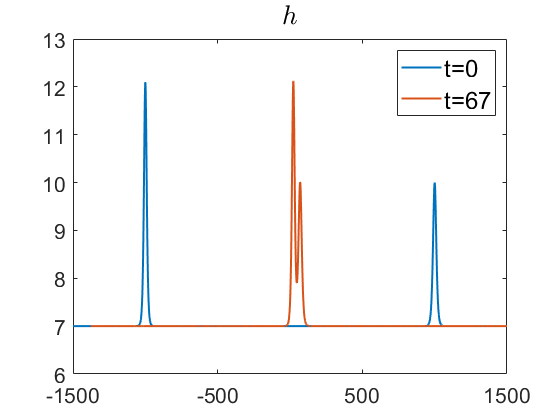}%\\
    \includegraphics[width=0.5\columnwidth%, height=9cm
    ]{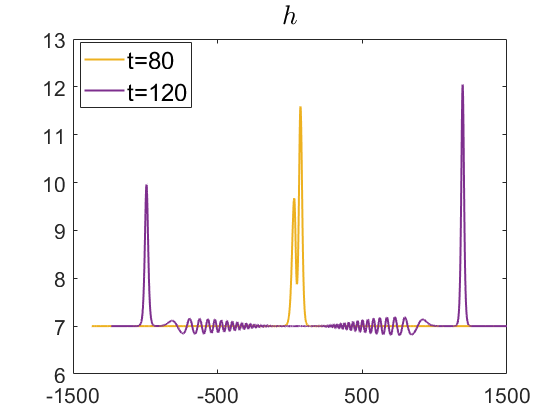}\\
    \includegraphics[width=0.5\columnwidth%, height=9cm
    ]{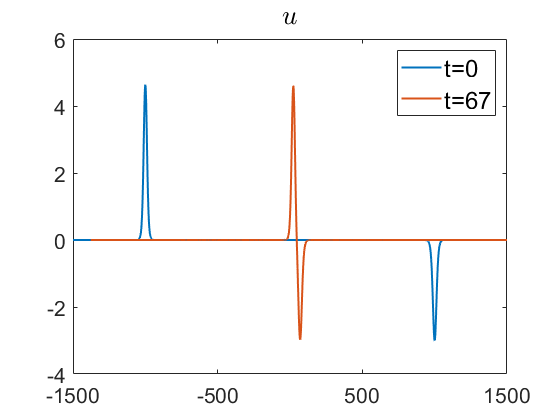}%\\
    \includegraphics[width=0.5\columnwidth%, height=9cm
    ]{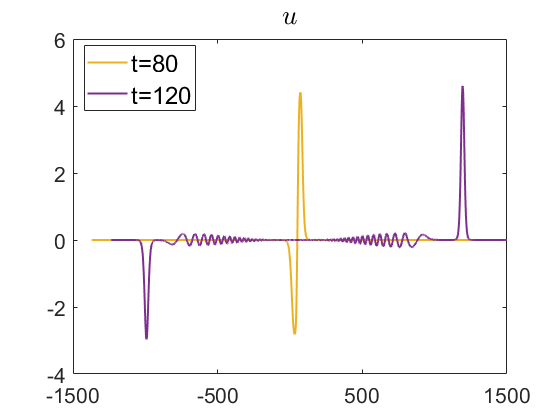}%\\
   \end{center}
    \caption{Snapshots of a collision of two counter-propagating solitons with maximum heights, respectively $h_{max} = 12.1, 10$ for two solitons at  $H =7$.  Top (bottom) row : height (velocity) field.
       \label{fig: hitting} 
        }
\end{figure}

\begin{figure}
    \begin{center}
   % \noindent\makebox[\columnwidth]{%
    \includegraphics[width=0.5\columnwidth, height=4cm
    ]{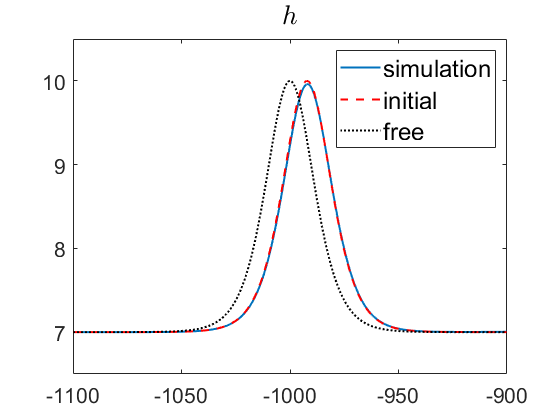}%\\
    \includegraphics[width=0.5\columnwidth, height=4cm
    ]{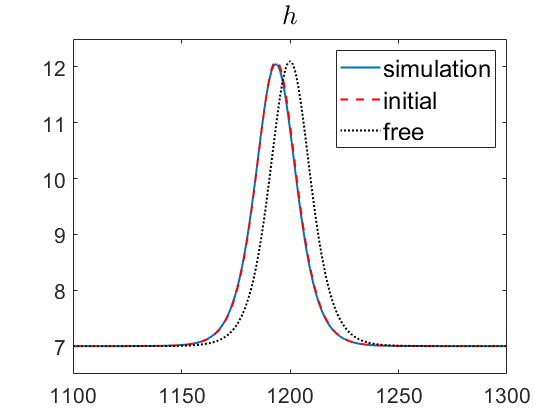}%\\
   \end{center}
    \caption{Difference between the positions and form of the left- moving (left) and right-moving (right) solitons after frontal collision of Fig. \ref{fig: hitting} (solid) with  their initial forms (dashed), and with the position each soliton would have if freely propagating (dotted).      
    \label{fig: comparisons2} 
        }
\end{figure}

We now test frontal collisions of two solitons. The results of one of such experiments are presented in Fig. \ref{fig: hitting}, and confirm, albeit with higher resolution, those obtained  earlier \citep{MirieSu, DutykhETAL, MitsotakisETAL}. As seen in the Figure, the counter-propagating solitons recover their form and steady motion, as follows from Fig. \ref{fig: comparisons2} but the collision produces pronounced finite-amplitude wave-tails which are left in the wake of each soliton. In this case the existence of the wake was also proven analytically  \citep{Byatt-Smith}. The energy of the wakes comes from diminishing soliton amplitudes, %as seen in Fig. \ref{fig: difference-hitting} 
and as a consequence of the speed, of the re-emerging solitons, which later gradually detach from the wake.  As in the case  of co-moving solitons, during the collision process the speed of the solitons momentarily changes, both soliton decelerating, as follows from Fig. \ref{fig: comparisons3}. 
%\begin{figure}
%    \begin{center}
   % \noindent\makebox[\columnwidth]{%
%    \includegraphics[width=0.5\columnwidth%, height=9cm
%    ]{2hittingwavesh_shifting.png}%\\
%    \includegraphics[width=0.5\columnwidth%, height=9cm
%    ]{2hittingwavesu_shifting.png}%\\
%   \end{center}
%    \caption{Comparison of the form of the $h$ (left) and $u$ (right) fields as follows from the simulation of the frontal collision of Fig. \ref{fig: hitting} (solid) with the initial form of each soliton (dashed).      
 %   \label{fig: comparisons2} 
 %       }
%\end{figure}
%\begin{figure}
%    \begin{center}
   % \noindent\makebox[\columnwidth]{%
 %   \includegraphics[width=0.5\columnwidth%, height=9cm
 %   ]{2hittingwavesh_differenceL.png}%\\
 %   \includegraphics[width=0.5\columnwidth%, height=9cm
 %   ]{2hittingwavesh_differenceR.png}%\\
 %  \end{center}
 %   \caption{Difference with the initial profiles of the  $h$-field for the left- (left) and right- (right) moving  solitons, as follows from the simulation of the frontal collision of Fig. \ref{fig: hitting}.      
 %   \label{fig: difference-hitting} 
%        }
%\end{figure}

\begin{figure}
    \begin{center}
   % \noindent\makebox[\columnwidth]{%
    \includegraphics[width=0.5\columnwidth, height=3cm
    ]{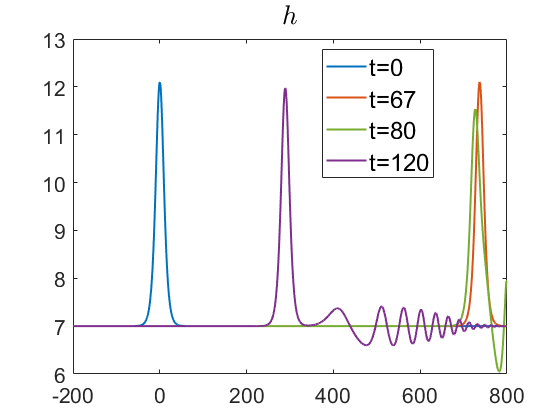}%\\
    \includegraphics[width=0.5\columnwidth, height=3cm
    ]{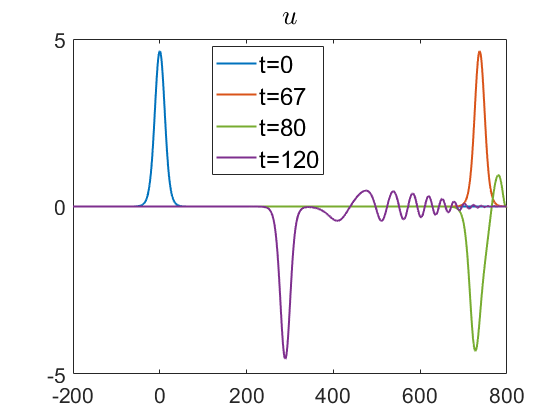}%\\
   \end{center}
    \caption{Snapshots of a collision of a soliton with a rigid wall situated at the right boundary of the simulation domain.   Left (right) panel: height (velocity) field.
       \label{fig: wall} 
        }
\end{figure}

A similar process is observed when a soliton hits a rigid wall: it recovers its form, but leaves a pronounced wave-tail in its wake after collision, see Fig. \ref{fig: wall}. In fact, a collision of two counter-propagating equal-height solitons is totally symmetric with respect to reflections $x \rightarrow -x$, if the origin is placed at the point of encounter, so collision with the wall represents its ``half'', as was already noticed in \citep{MirieSu}. 
\subsection{Emergence of multi-soliton configurations from localized initial conditions}
\begin{figure}
    \begin{center}
   % \noindent\makebox[\columnwidth]{%
    \includegraphics[width=\columnwidth, height=4cm
    ]{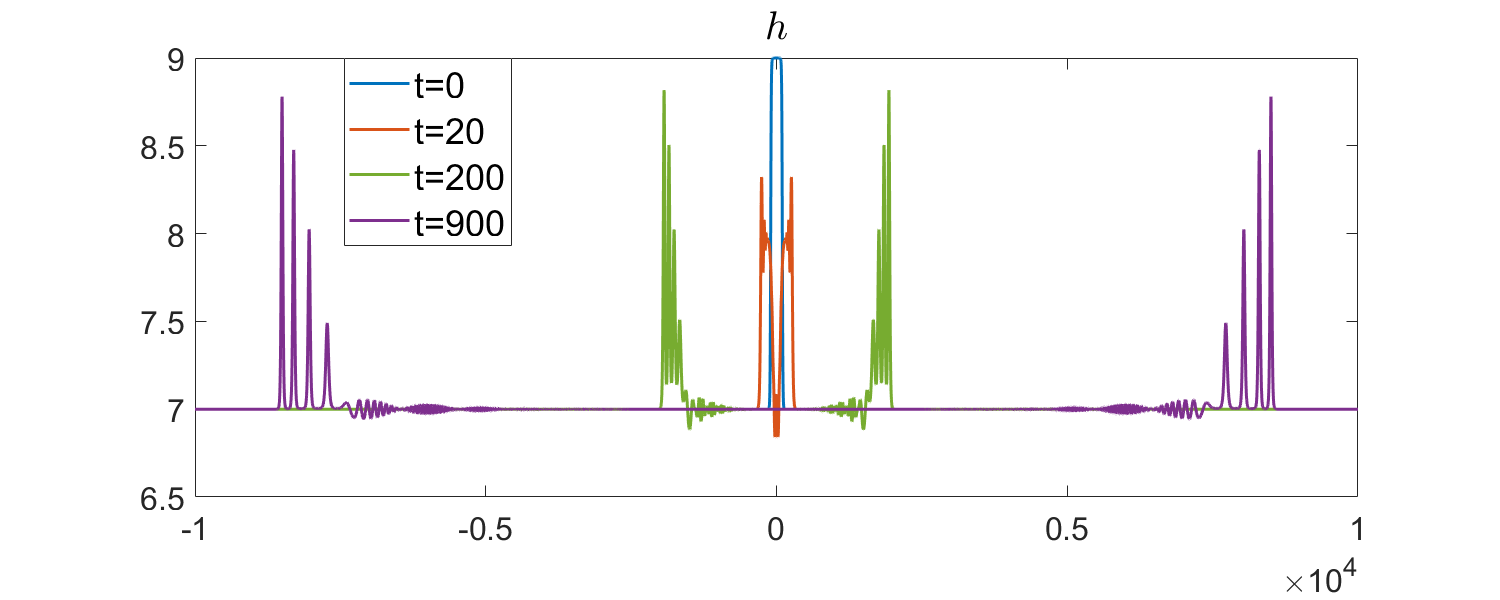}\\
    \includegraphics[width=\columnwidth, height=4cm
    ]{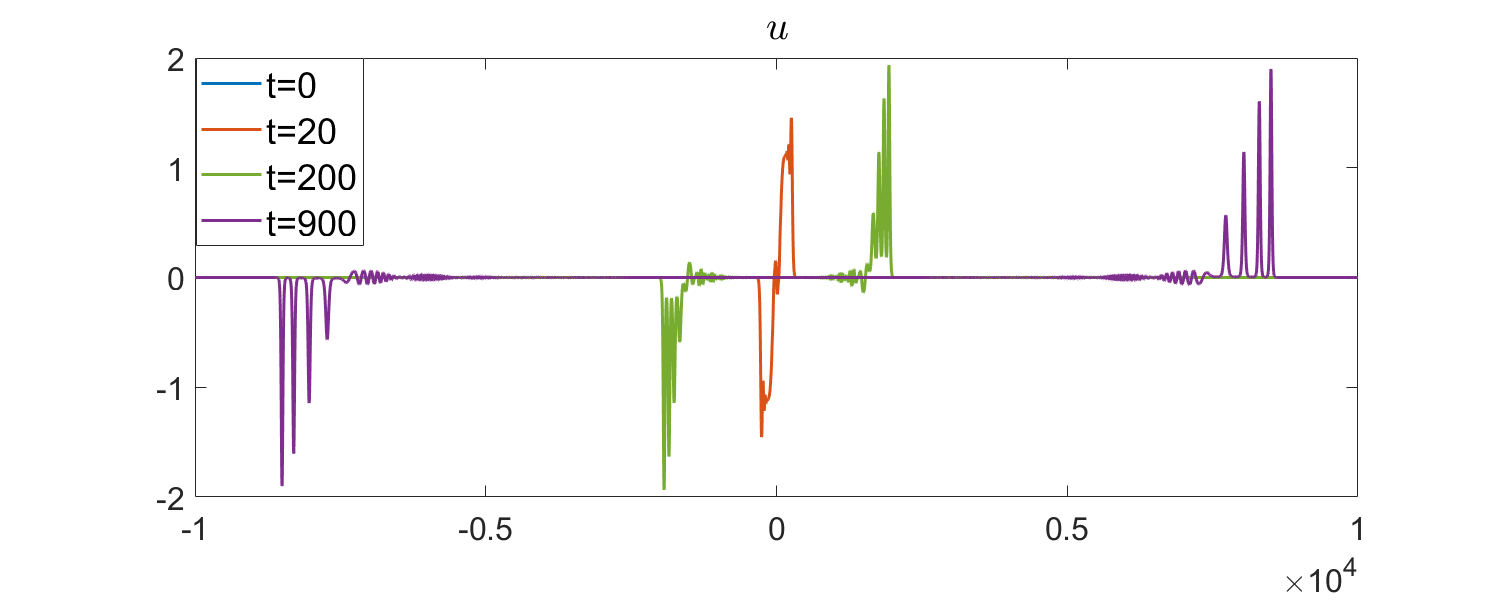}%\\
   \end{center}
    \caption{Snapshots of the evolution of an initial ``flat-bump'' distribution of $h$ obtained by superposition of two $\tanh$ profiles of opposite signs, with $u \equiv 0$.    Top (bottom) panel: height (velocity) field.
       \label{fig: initial} 
        }
\end{figure}

We finally check whether  localized initial conditions give rise a sequence of solitons, like it is the case for KdV solitons. We tested different  initial distributions of $h$, with zero initial $u$. The overall result is that multi-soliton configurations do emerge, accompanied by a small-amplitude wave tails, but the number of solitons in each direction is conditioned by the shape and amplitude of the initial perturbation. We present an example of a simulation with a symmetric ``flat-bump'' producing sequences of 4 solitons at each of its sides in Fig. \ref{fig: initial}. Remarkably, the wave wake behind the soliton system is pulsating in this particular case. For example, for an initial Gaussian disturbance,  by increasing its amplitude we observed  from one to four solitons to emerge (not shown).  
%\begin{figure}
%    \centering
%    \noindent\makebox[\columnwidth]{%
%    \includegraphics[width=\columnwidth]{}}
%    \caption{ }
%    \label{}
%\end{figure}
\subsection{Interaction of solitons with topographic obstacles}
\begin{figure}
    \begin{center}
   % \noindent\makebox[\columnwidth]{%
    \includegraphics[width=0.5\columnwidth, height=4cm
    ]{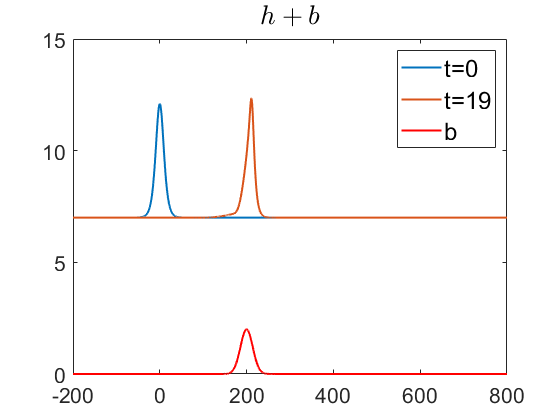}%\\
    \includegraphics[width=0.5\columnwidth, height=4cm
    ]{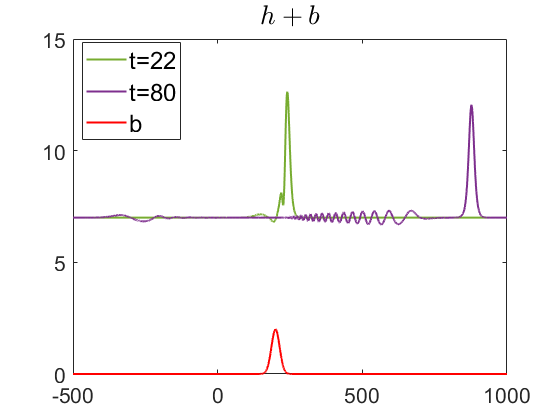}\\
    \includegraphics[width=0.5\columnwidth, height=4cm
    ]{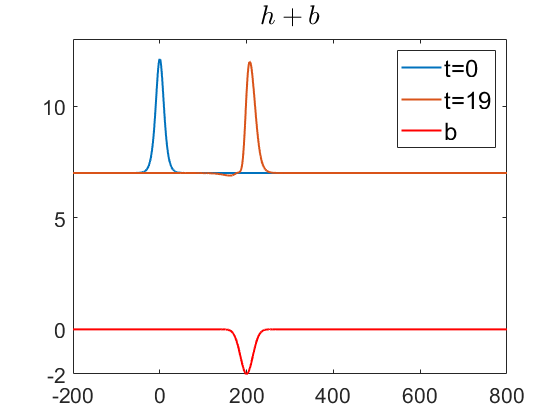}%\\
    \includegraphics[width=0.5\columnwidth, height=4cm
    ]{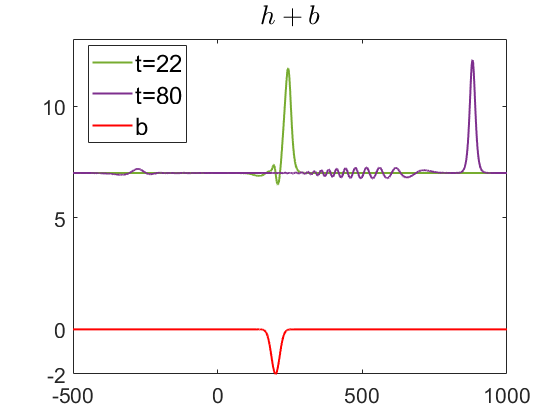}%\\
   \end{center}
    \caption{Snapshots of a height field of a  soliton hitting a localized bump (top row) and a localized dip (bottom row).
       \label{fig: bumps} 
        }
\end{figure}

\begin{figure}
    \begin{center}
   % \noindent\makebox[\columnwidth]{%
    \includegraphics[width=0.5\columnwidth, height=4cm
    ]{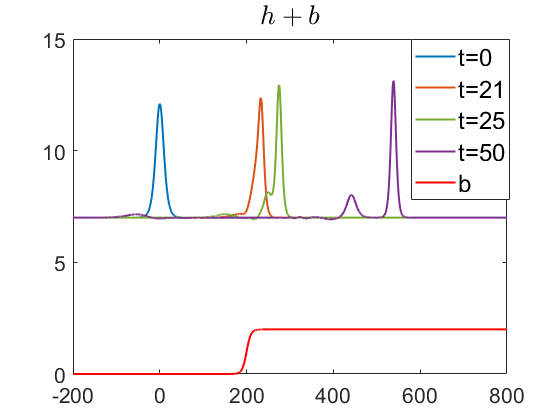}%\\
    \includegraphics[width=0.5\columnwidth, height=4cm
    ]{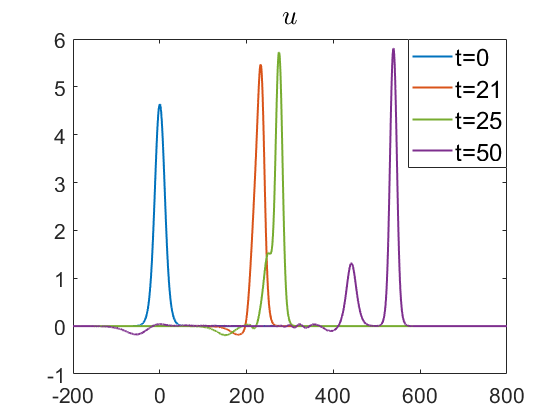}\\
    \includegraphics[width=0.5\columnwidth, height=4cm
    ]{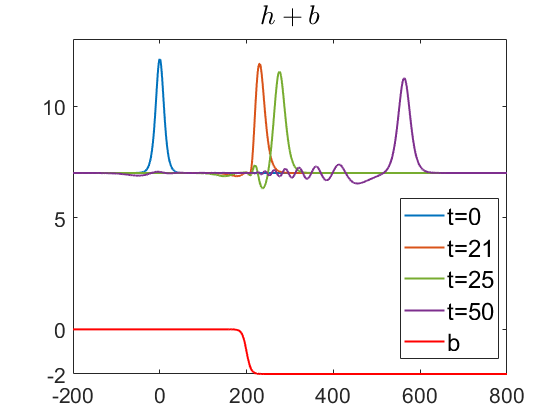}%\\
    \includegraphics[width=0.5\columnwidth, height=4cm
    ]{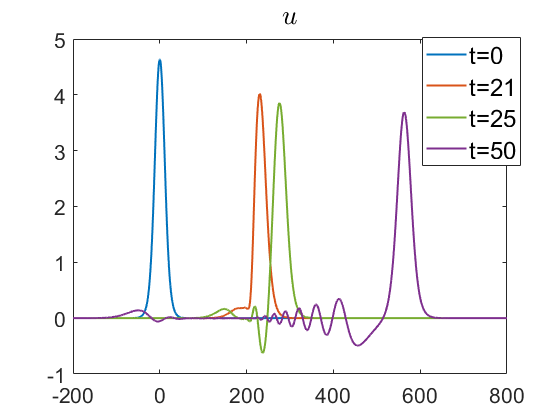}%\\
   \end{center}
    \caption{Snapshots of a height (left column) and velocity (right column) fields of a  soliton hitting a rising (top row) and a descending  (bottom row) escarpment of $\tanh$ shape.
       \label{fig: escarpments} 
        }
\end{figure}

Next, we investigated interactions of solitons with topographic obstacles: localized bumps, or dips, and escarpments. The results are presented in Figs. \ref{fig: bumps}, \ref{fig: escarpments} and show that soliton keeps its coherence passing over the first two, leaving 
only a detaching wave-train behind, while mounting an escarpment it generates a trailing smaller-amplitude soliton, a behavior observed in laboratory, and early numerical experiments by \cite{Seabra-SantosETAL}. This is not the case of descending escarpment, where only a wave-wake is produced. These results can be qualitatively understood in view of those on initial-value problem. As follows from \eqref{2.2}, the amplitude and velocity of the soliton increase with diminishing $h_{\infty}$. At the same time, the velocity of small-amplitude waves decreases. So mounting an escarpment the soliton should accelerate to keep this form, which is energetically impossible, so it adjusts, according to the previous results, producing a two-soliton configuration. By the same reason, descending escarpments leads to deceleration, which can be achieved by wave emission.

\section{Discussion}
\label{sec: disc}
Our simulations thus exhibit a striking resemblance in the behavior of SGN solitons with their analogs in the KdV system. This could be expected in the weak-nonlinearity ($\equiv$ small-amplitude) limit in view of the fact that KdV is an asymptotic limit of the SGN, but is surprising for amplitudes in a range $\mathcal{O}(1) - \mathcal{O}(10)$.  First of all, multi-soliton configurations ubiquitously arise from localized initial conditions. The collisions of co-moving solitons exhibit the same scenarios as in the KdV case, completely recovering after multiple (simultaneous or sequential) collisions, modulo small-amplitude wave-tails they leave in their wake.  While such behavior was already observed in the literature for pairwise collisions, the  extension to  triple collisions we obtained strongly enhances similarity with  the KdV system.  Counter-propagating  solitons also exit collisions recovering their form, but leaving  a more pronounced wave wake behind.   Moreover, the solitons retain their shape, emitting a wave-wake,  while hitting localized topographic bumps or dips, and descending escarpments. At rising escarpments the solitons produce secondary trailing ones.  

Thus, our results demonstrate an attracting character of multi-soliton configurations, and their surprising stability. We can not claim, of course, that the SGN system is completely integrable, like its descendant, the KdV equation, but ubiquity and  high regularity of its multi-soliton solutions revealed by the simulations is quite remarkable.  
\bibliographystyle{jfm}
%\bibliography{bib_SGN}
\bibliography{bib_SGN_v1-a1}

\end{document}